\documentclass[english,final, aps]{revtex4}
\usepackage[T1]{fontenc}
\usepackage[latin9]{inputenc}
\usepackage{amsmath}
\usepackage{amssymb}
\usepackage{graphicx}

\makeatletter

\providecommand{\tabularnewline}{\\}

\@ifundefined{textcolor}{}
{%
 \definecolor{BLACK}{gray}{0}
 \definecolor{WHITE}{gray}{1}
 \definecolor{RED}{rgb}{1,0,0}
 \definecolor{GREEN}{rgb}{0,1,0}
 \definecolor{BLUE}{rgb}{0,0,1}
 \definecolor{CYAN}{cmyk}{1,0,0,0}
 \definecolor{MAGENTA}{cmyk}{0,1,0,0}
 \definecolor{YELLOW}{cmyk}{0,0,1,0}
}

\makeatother

\usepackage{babel}

\begin{document}

\title{Search for Physics beyond Standard Model at the Precision Frontiers}

\author{A. Aleksejevs and S. Wu}

\address{Grenfell campus of Memorial University, Corner Brook, Canada}

\author{S. Barkanova}

\address{Acadia University, Wolfville, Canada}

\author{Y. Bystritskiy}

\address{Joint Institute for Nuclear Research, Dubna, Russia}

\author{V. Zykunov}

\address{Belarussian State University of Transport, Gomel, Belarus}

\begin{abstract}
The article outlines the recent developments in the theoretical and computational approaches to the higher-order
electroweak effects needed for the accurate interpretation of MOLLER and Belle II experimental
data, and shows how new-physics particles enter at the one-loop level.
By analyzing the effects of $Z'$-boson on the polarization asymmetry, we show how this hypothetical interaction
carrier may influence the future experimental results.
\end{abstract}
\maketitle

\section{Introduction}

The recent availability of computer-algebra tools in particle
physics research provides an unique opportunity to perform the Next-to-Leading-Order
(NLO) and Next-to-NLO (NNLO) Standard Model (SM) calculations with
the high degree of precision required by MOLLER and Belle
II. Here, full SM calculations are required, with no
approximations at the NLO level, and include leading order NNLO contributions, which can only be achieved with some degree of automatization. We do this
for both MOLLER ($e^{-}+e^{-}\rightarrow e^{-}+e^{-}$) and Belle
II ($e^{+}+e^{-}\rightarrow\mu^{+}+\mu^{-}$), and compare the results
of calculations performed with the different sets of renormalization conditions
using the on-shell renormalization. That provides a straightforward test
of gauge invariance for the polarization asymmetry. A discrepancy between SM predictions and experimental
measurements would signal the physics beyond the SM. Since MOLLER and
Belle II are the most sensitive to the parity-violating (PV) interaction,
we include U(1)$'$ extension of SM with a mass mixing scenario, which
results in extension of SM by the parity-violating $Z'$-boson. Our analysis
for $Z'$ extends to NLO level giving us a refined set of constrains
on the coupling and mass. First, we start with details on NLO and NNLO (quadratic) calculations for
MOLLER and then continue with Belle II. In the second part of the paper, we provide
results and analysis of the polarization asymmetry with $Z'$-boson present
at LO and NLO orders.

\section{SM Predictions for Polarization Asymmetry in MOLLER and
Belle II}

We consider two processes, $e_{k_{1}}^{-}+e_{k_{2}}^{-}\rightarrow e_{k_{3}}^{-}+e_{k_{4}}^{-}$, for MOLLER, 
and $e_{k_{1}}^{+}+e_{k_{2}}^{-}\rightarrow\mu_{k_{3}}^{+}+\mu_{k_{4}}^{-}$, for Belle II. 
For MOLLER, the most sensitive observable
to PV new physics (aka $Z'$) is the polarization asymmetry:
%
%
\begin{eqnarray}
A_{LR} & = & \frac{\sigma_{L}-\sigma_{R}}{\sigma_{L}+\sigma_{R}}\simeq\frac{2\Re(M_{\gamma}M_{Z}^{+}+M_{\gamma}M_{Z'}^{+}+M_{Z}M_{Z'}^{+})_{LR}}{\sigma_{L}+\sigma_{R}}.\label{eq:2}
\end{eqnarray}
In Eq.\ref{eq:2}, $Z'$ will enter numerator of asymmetry through the interference
term. For the $e^{-}+e^{-}\rightarrow e^{-}+e^{-}$ process, the asymmetry at
LO order given by the following expression:
\begin{eqnarray}
A_{LR{\rm(MOLLER)}}^{0} & = & \frac{s}{m_{W}^{2}}\frac{y(1-y)}{1+y^{4}+(1-y)^{4}}\frac{1-4s_{W}^{2}}{s_{W}^{2}}.\label{eq:3}
\end{eqnarray}
Here, $y=-t/s$, 
the set of Mandelstam variables is used: $s=(k_{1}+k_{2})^{2}$, $t=(k_{3}-k_{1})^{2}$ and $u=(k_{1}-k_{4})^{2}$,  
and the sine of Weinberg mixing angle is denoted as $s_{W} \equiv\sin \theta_{W}$. 
As one can see, the LO asymmetry is proportional to $1-4s_{W}^{2}$, which results
in strong sensitivity to $s_{W}^{2}$. This provides an excellent
opportunity for the precision measurements of $s_{W}^{2}$, or, accordingly, the measurement of the weak charge of electron. Although PV asymmetry
in Eq.\ref{eq:3} is quite small, the accuracy of modern experiments
exceed the accuracy of the theoretical result at LO order; 
the NLO order calculations have been completed by number of authors
\cite{key-1,key-2,key-3}. Generally, we can express perturbative expansion
(up to NNLO) for differential scattering cross-section in orders of $\alpha$ as:
\begin{eqnarray}
\frac{d\sigma}{d\cos\theta} & = & \frac{\pi^{3}}{2s}\Big|M_{0}+M_{1}+M_{2}\Big|^{2}\simeq\nonumber \\
\nonumber \\
& &\frac{\pi^{3}}{2s}
\left( \alpha^{2}M{}_{0}^{'}M{}_{0}^{'+}
+\alpha^{3}2\Re M{}_{0}^{'}M{}_{1}^{'+}
+\alpha^{4} (M_{1}^{'}M_{1}^{'+}+2\Re M{}_{0}^{'}M{}_{2}^{'+}) \right),\label{eq:4}
\end{eqnarray}
where matrix elements $M_{i}$ are related to $M_{i}^{'}$
by $M_{i}=\alpha^{i+1}M_{i}^{'}$. The first term corresponds
to LO, the second to NLO and the third forms NNLO contribution, which comprises
from quadratic term ($\alpha^{4}M{}_{1}^{'}M{}_{1}^{'+}$) and two-loops
($\alpha^{4}2\Re M{}_{0}^{'}M{}_{2}^{'+}$) contribution. The NLO contribution to LO asymmetry is rather big ($\sim 69\%$)
\cite{key-3}, and in order to match 1\% MOLLER uncertainty, we calculated a full set NNLO (quadratic) \cite{key-4} 
and leading order NNLO (two-loops) contributions \cite{key-5} (and references therein).
The precision is essential, so we control it in two ways. 
First, we applied ``on paper''
on-shell calculations using renormalization conditions of \cite{key-10}
and low energy approximations $\frac{r}{m_{Z,W}^{2}} \ll 1$ (here $r=s,|t|,|u|$).
Second, we performed semi-automated \cite{key-11,key-12,key-13,key-14,key-15}
calculations for the full set of Feynman diagrams without any approximations
and using renormalization conditions of \cite{key-16}. This approach
was implemented for NLO and NNLO (quadratic) contributions. The semi-automated full two-loop calculations are yet to be completed
which is our next goal. Let us demonstrate how these two approaches
compare to each other. First, we introduce a correction to the asymmetry, as:
\begin{eqnarray}
\delta_{A}^{C} & = & \frac{A_{LR}^{C}-A_{LR}^{0}}{A_{LR}^{0}},\label{eq:5}
\end{eqnarray}
where $A_{LR}^{C}$ stands for the NLO-corrected asymmetry.

\begin{figure}
\begin{centering}
\includegraphics[scale=0.38]{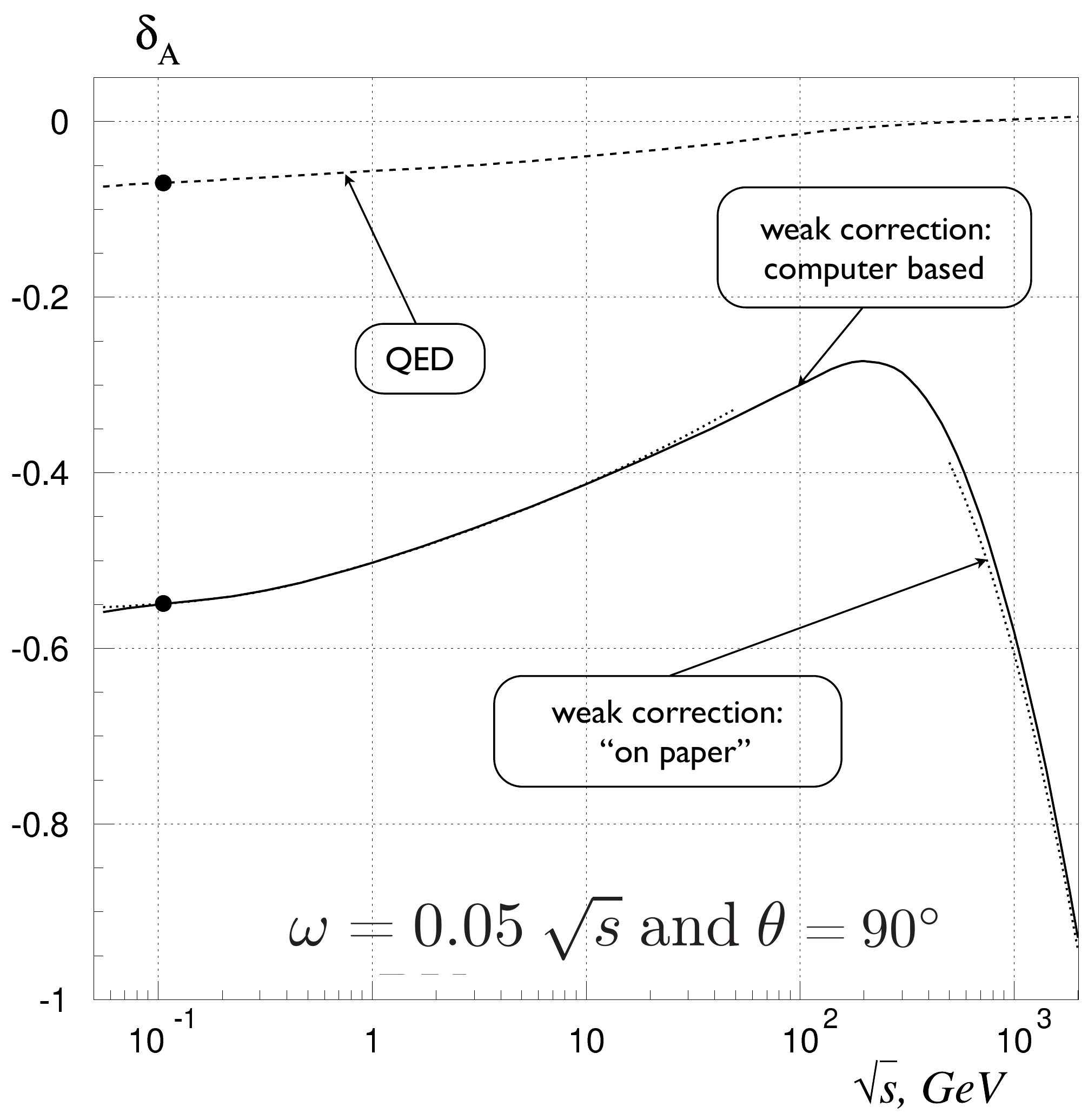} \includegraphics[scale=0.41]{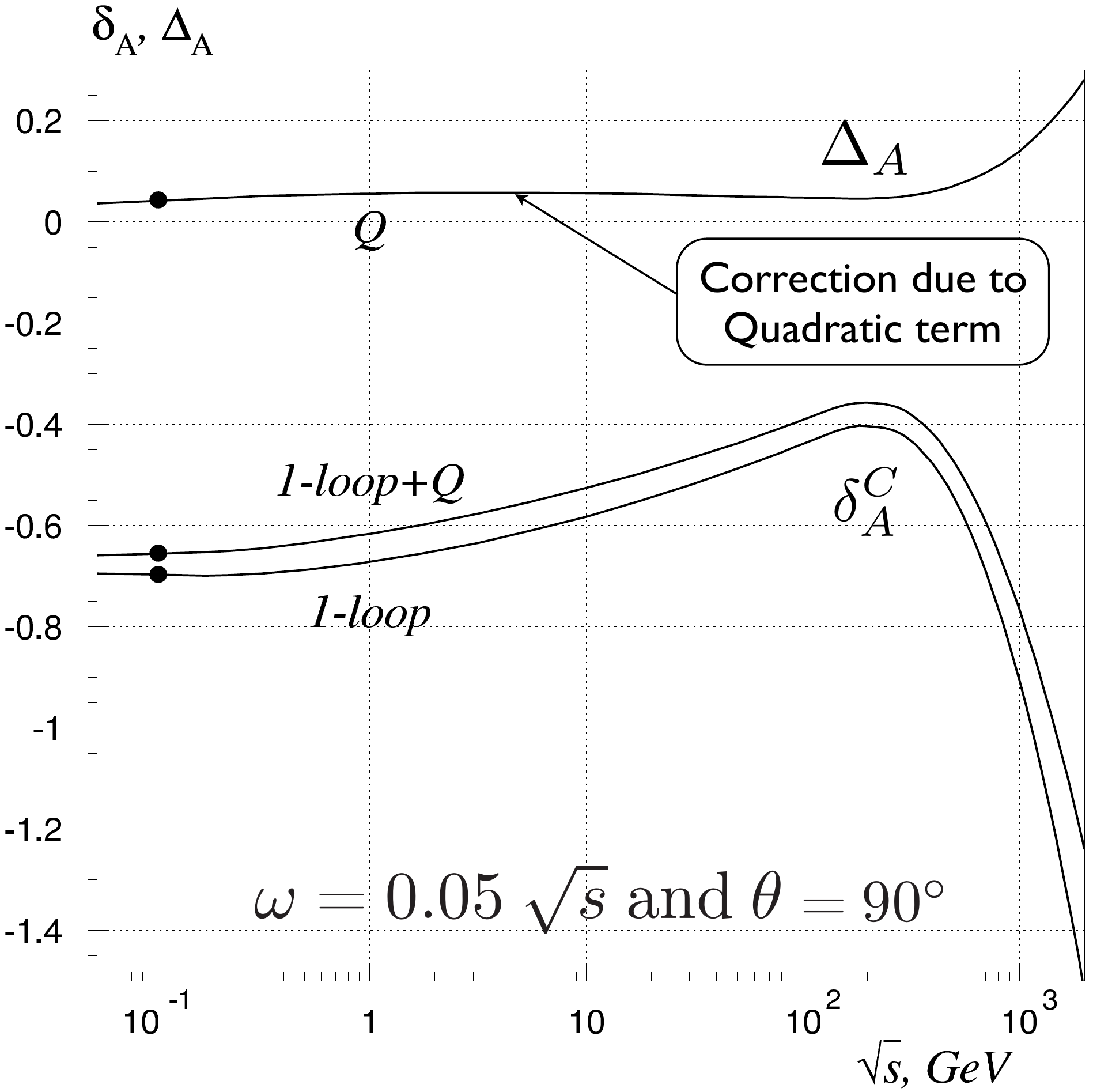}
\par\end{centering}

\caption{The graphs show corrections to the polarization asymmetry
at the $\theta=90^\circ$ and soft photon cut $\omega=0.05\sqrt{s}$.
In the left graph, dashed line shows correction due to QED effects,
solid line represents weak correction based on semi-automated calculations
and dotted line gives ``on-paper'' approximated calculations for
the same correction. Right graph shows correction for pure quadratic
(``Q'') contribution ($\Delta_{A}$), NLO (``1-loop'') and NLO+NNLO
(``1-loop-Q'') results. Black dot on both graphs corresponds to
the $E_{\rm lab}=11\ {\rm GeV}$.}

\label{fig1}
\end{figure}

\begin{figure}
\begin{centering}
\includegraphics[scale=0.30]{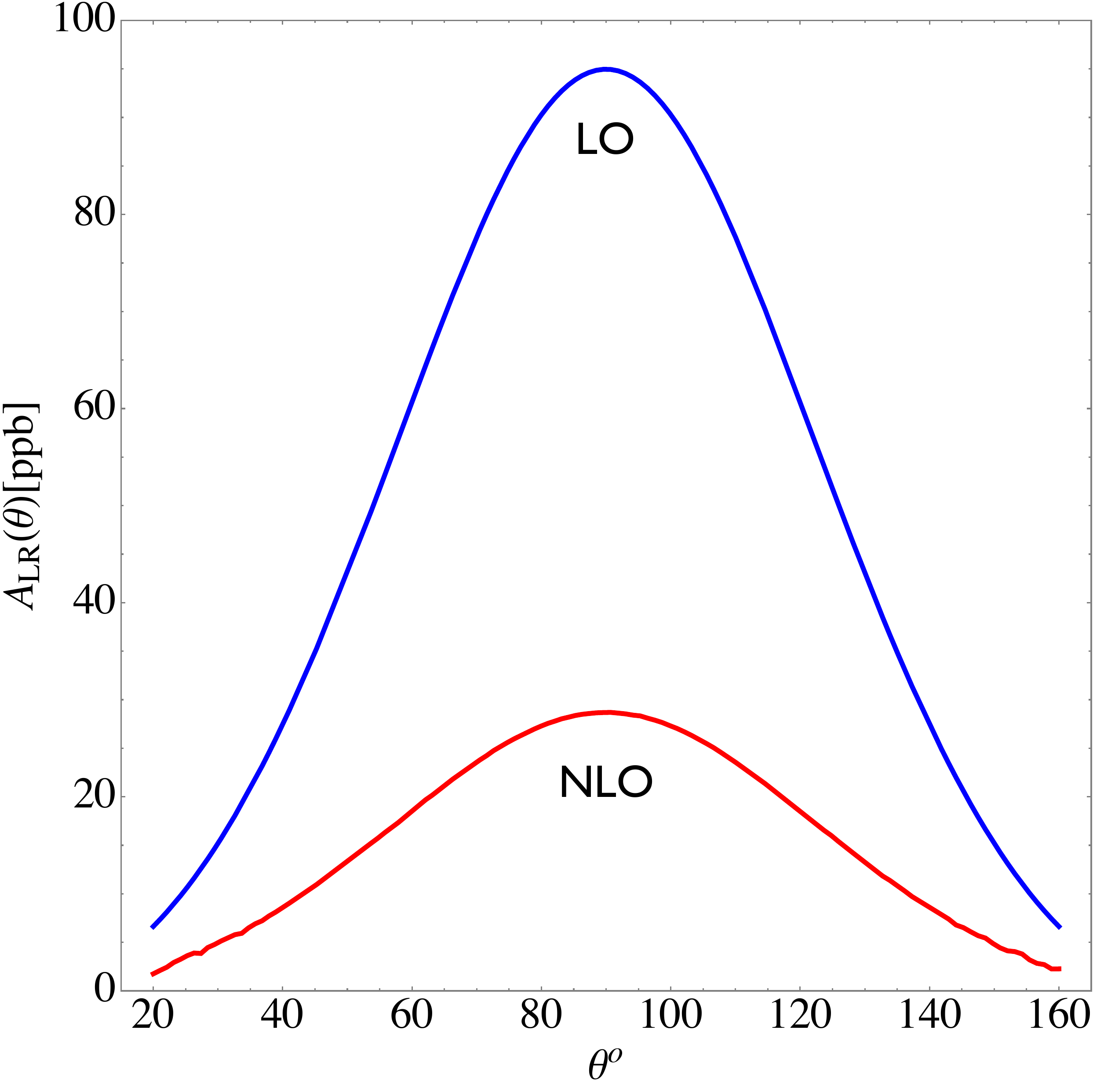} \includegraphics[scale=0.31]{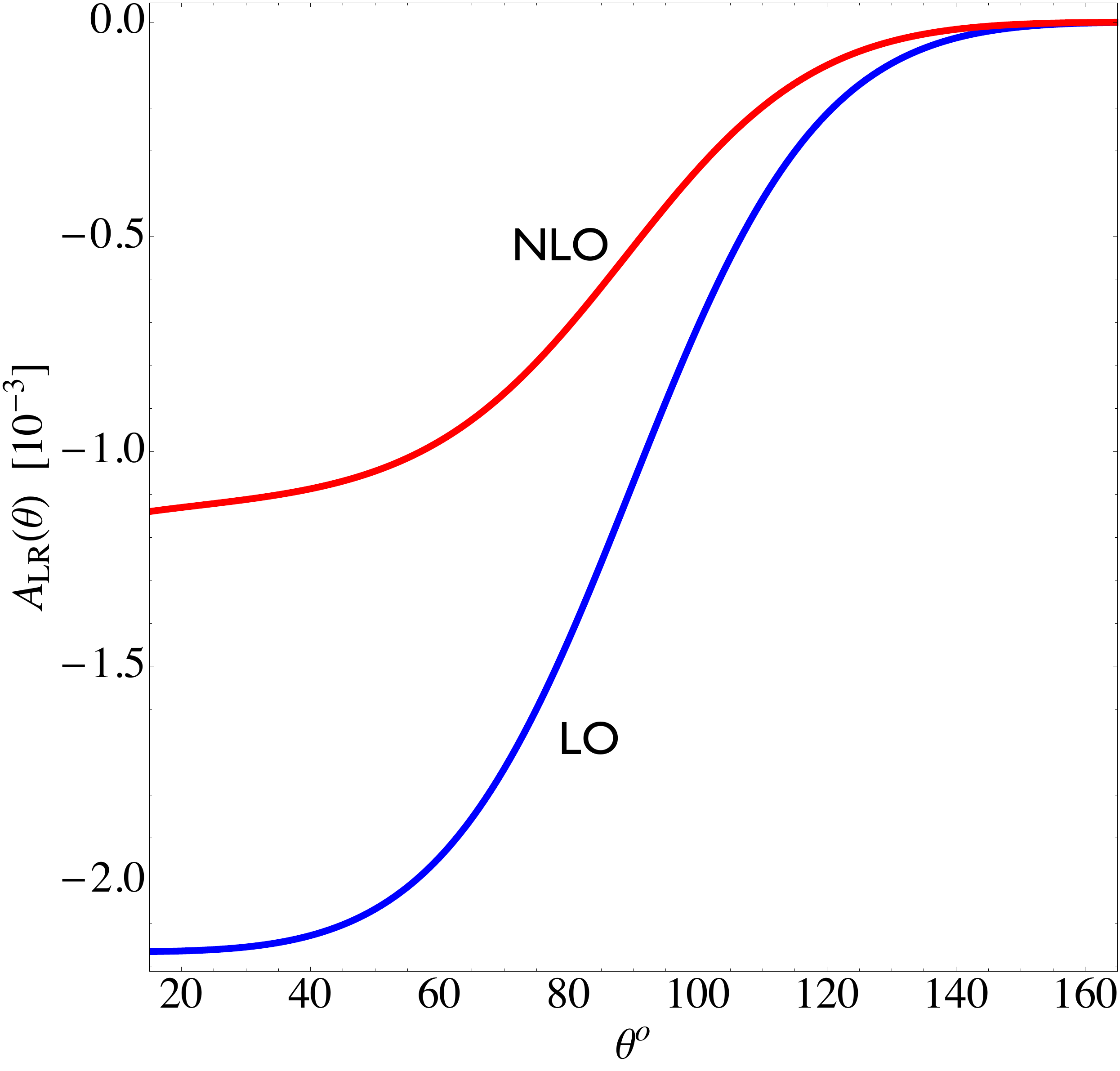}
\par\end{centering}

\caption{LO and NLO corrected polarization asymmetries for the MOLLER (left,
$E_{\rm lab}=11\ {\rm GeV}$) and Belle II (right, $\sqrt{s}=10.57\ {\rm GeV}$)
experiments for the $20^\circ < \theta<160^\circ $ in the CM reference
frame. }

\label{fig2}
\end{figure}

If we take $\alpha=1/137.0359$, $m_{W}=80.398\ {\rm GeV}$, $m_{Z}=91.1876\ {\rm GeV}$
and kinematics relevant to MOLLER experiment ($E_{\rm lab}=11\ {\rm GeV}$),
we can see in Fig.\ref{fig1} (right plot) that results obtained in
both approaches differ less than 0.1\%. We find that the NNLO (quadratic)
contribution, Fig.\ref{fig1} (left plot), 
\begin{eqnarray}
\Delta_{A} & = & \frac{A_{LR}^{\rm 1-loop-Q}-A_{LR}^{\rm 1-loop}}{A_{LR}^{0}},\label{eq:5a}
\end{eqnarray}
is responsible for $\sim 5\%$ suppression of the total correction
at $E_{\rm lab}=11\ {\rm GeV}$. This is a clear signal that, in the light of proposed precision experiments, the NNLO contributions
are very important. Similar to the Moller process, $e^{+}+e^{-}\rightarrow\mu^{+}+\mu^{-}$
polarization asymmetry (first addressed in \cite{key-17}), also shows strong sensitivity to the $s_{W}^{2}$:
\begin{eqnarray}
A_{LR({\rm Belle\,II})}^{0} & = & -\frac{s}{4m_{W}^{2}}\frac{(y-1)^{2}}{2(y-1)y+1}\frac{1-4s_{W}^{2}}{s_{W}^{2}}.\label{eq:6}
\end{eqnarray}
We improve the precision by implementing the same two-way approach, for Belle II kinematics specifically, 
taking into account the full set of NLO electroweak corrections. For $\sqrt{s}=10.57\ {\rm GeV}$,
Table \ref{table1} shows our results for the NLO relative correction to
unpolarized cross section ($\delta_{00}^{C}=\frac{\sigma_{00}^{C}-\sigma_{00}^{0}}{\sigma_{00}^{0}}$)
computed using the semi-automated (SA) and ``on-paper'' calculation methods,
in the on-shell renormalization. 
\begin{table}
\begin{centering}
\begin{tabular}{|c|c|c|c|c|c|c|c|c|c|}
\hline 
$\theta,^\circ$ & 10 & 30 & 50 & 70 & 90 & 110 & 130 & 150 & 170\tabularnewline
\hline 
\hline 
``on-paper'' & 0.0180 & $-0.0456$ & $-0.0738$ & $-0.0935$ & $-0.1099$ & $-0.1264$ & $-0.1460$ & $-0.1743$ & $-0.2378$ \tabularnewline
\hline 
SA & 0.0179 & $-0.0455$ & $-0.0738$ & $-0.0934$ & $-0.1099$ & $-0.1263$ & $-0.1459$ & $-0.1742$ & $-0.2372$ \tabularnewline
\hline 
\end{tabular}
\par\end{centering}

\caption{Relative NLO correction to unpolarized cross section $\delta_{00}^{C}$
in $e^{+}+e^{-}\rightarrow\mu^{+}+\mu^{-}$ process for the various 
angles in CM reference frame. Both approaches used soft-photon
approximation in the treatment of infrared divergencies. Soft photon
cut is $\omega=0.05\sqrt{s}$ .}

\label{table1}
\end{table}
Evidently, difference in both approaches, for the broad range of scattering
angles, is negligible. In Fig.\ref{fig2} we show the results
for the NLO corrected asymmetries for both MOLLER and Belle II experiments.

Although the NLO contribution (Fig.\ref{fig2}) for both processes
is significant, judging by excellent agreement between two approaches
presented here, the theoretical uncertainty at NLO level is at sub-percent
level. However, that result does not include the NNLO contribution, needed to
interpret ultra-precision measurements.

\section{Beyond the Standard Model Physics with Dark Vector}

\begin{figure}
\begin{centering}
\includegraphics[scale=0.44]{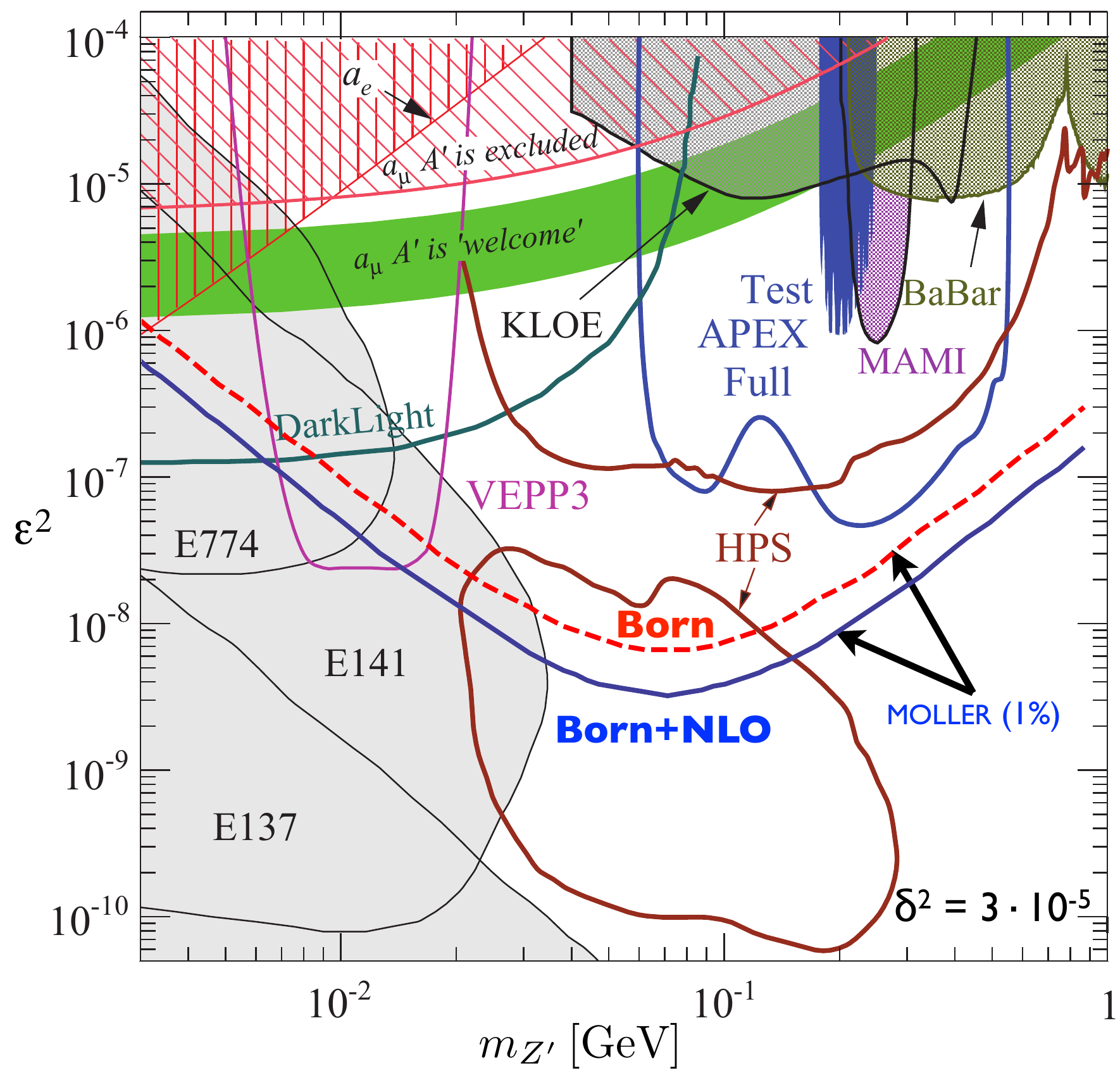} \includegraphics[scale=0.287]{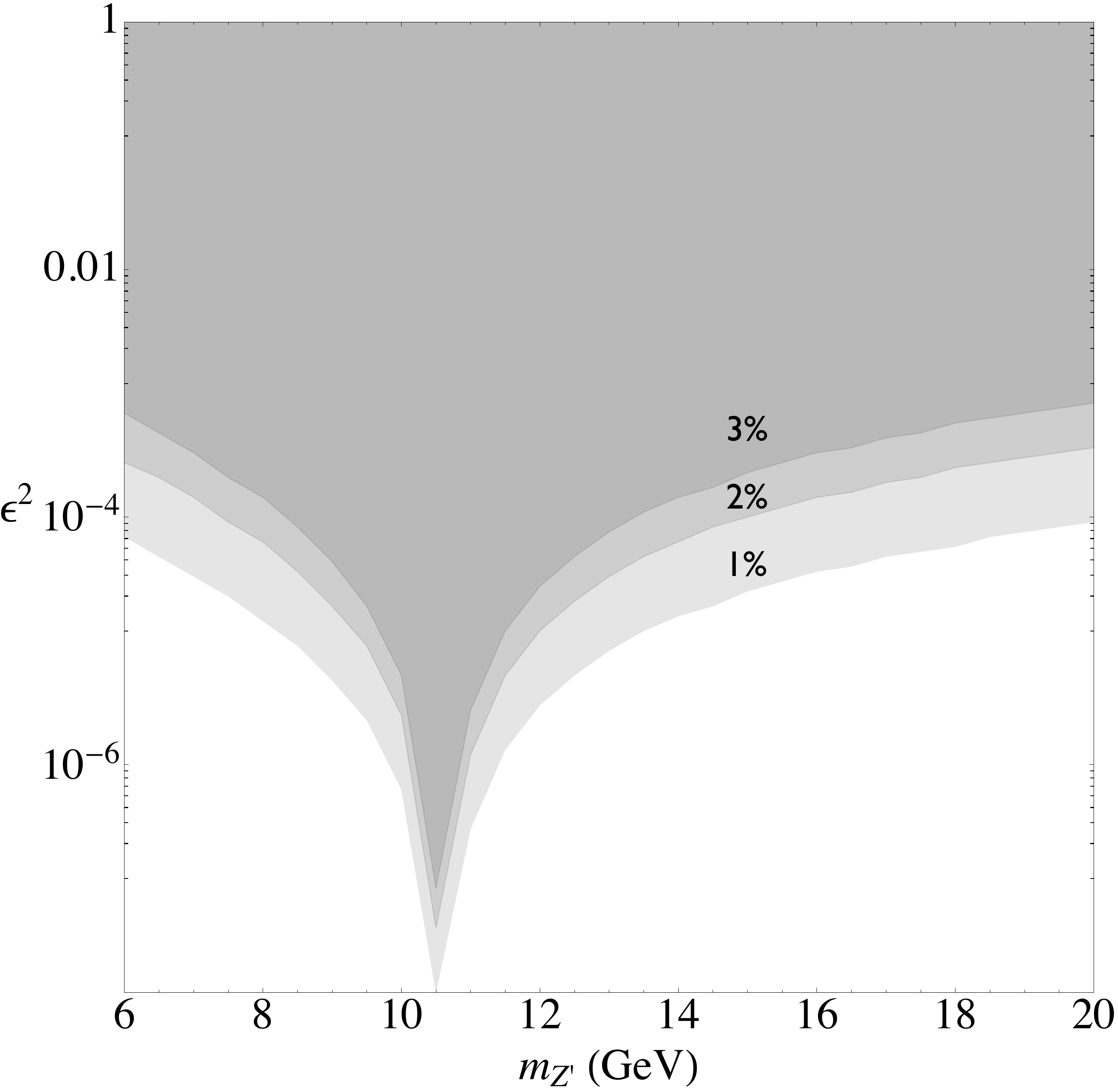}
\par\end{centering}

\caption{Exclusion plots for MOLLER (left) and Belle II (right). For MOLLER,
we show exclusion for 1\% deviation from SM prediction. Here, $Z'_{\mu}$
is included either at LO (Born) or up to NLO (Born+NLO) orders. Plot
for Belle II shows up-to-NLO exclusion regions for 1\%, 2\% and 3\%
deviations from SM prediction. For both plots, we take $\delta^{2}=3\cdot10^{-5}$.}

\label{fig3}
\end{figure}

The discrepancy between experimental
data and SM theoretical predictions would signal new interaction
carriers. We use the simple U(1)$'$ extension
of SM proposed in \cite{key-18}, which
uses kinetic type of mixing between dark vector ($A'_{\mu}$) and
hyper-charge ($B_{\mu}$) fields:
\begin{eqnarray}
\mathfrak{L}_{kin} & = & -\frac{1}{4}B_{\mu\nu}B^{\mu\nu}+\frac{1}{2}\frac{\epsilon}{\cos\theta_{W}}B_{\mu\nu}A'^{\mu\nu}-\frac{1}{4}A'_{\mu\nu}A'^{\mu\nu},\label{eq:7}
\end{eqnarray}
where fields tensors are given by $\{A'_{\mu\nu},\,B_{\mu\nu}\}=\partial_{\mu}\{A'_{\nu},\,B_{\nu}\}-\partial_{\nu}\{A'_{\mu},\,B_{\mu}\}$,
with $B_{\mu}=\cos\theta_{W}A_{\mu}-\sin\theta_{W}Z_{\mu}$ and $\epsilon$
is the $(B_{\mu}-A'_{\mu})$ kinetic mixing parameter. With SM Higgs doublet plus the Higgs singlet 
(used for breaking the U(1)$'$ symmetry and giving mass to $A'_{\mu}$), a Lagrangian describing
interaction between the SM fermions and the dark vector boson $A'_{\mu}$ is:
\begin{flalign}
\mathfrak{L}_{int}= & \,-eQ_{f}\bar{f}\gamma^{\mu}f\cdot(V{}_{\mu}+\epsilon A'_{\mu})-\nonumber \\
\nonumber \\
 & \frac{e}{\sin\theta_{W}\cos\theta_{W}}\bar{f}(c_{V}^{f}\gamma^{\mu}+c_{A}^{f}\gamma^{\mu}\gamma_{5})f
\cdot(Z{}_{\mu}+\epsilon_{Z'}A'_{\mu}).\label{eq:2-1}
\end{flalign}
Here, $Q_{f}$ is the charge of the fermion in units of $e$, and the
$c_{V}^{f}$ and $c_{A}^{f}$ constants are usual SM vector and axial-vector
coupling strengths, respectively. As we can see from Eq.\ref{eq:2-1},
the $A'_{\mu}$ couples to fermions through both parity-conserving
and violating terms, which is similar to the weak $Z_{\mu}$ coupling.
That type of the $A'_{\mu}$ in \cite{key-19} is called the
dark $Z_{\mu}'$-boson and derived from an additional mass mixing
term characterized by mixing parameter $\epsilon_{Z'}=\frac{m_{Z'}}{m_{Z}}\delta$.
Here, $m_{Z'}$ is the the mass of the dark $Z_{\mu}'$-boson and
$\delta$ is an arbitrary model-dependent parameter. The fact what
$Z'_{\mu}$ is represented as a superposition of mixings between dark
vector with electromagnetic and $Z$-boson fields makes it possible to
include $Z'_{\mu}$ at NLO level. We include $Z'_{\mu}$ at NLO
 in order to match NLO calculations for our SM predictions. Our
results are shown in the form of exclusion plots in Fig.\ref{fig3} for MOLLER
(left) and Belle II (right).

\begin{figure}
\begin{centering}
\includegraphics[scale=0.207]{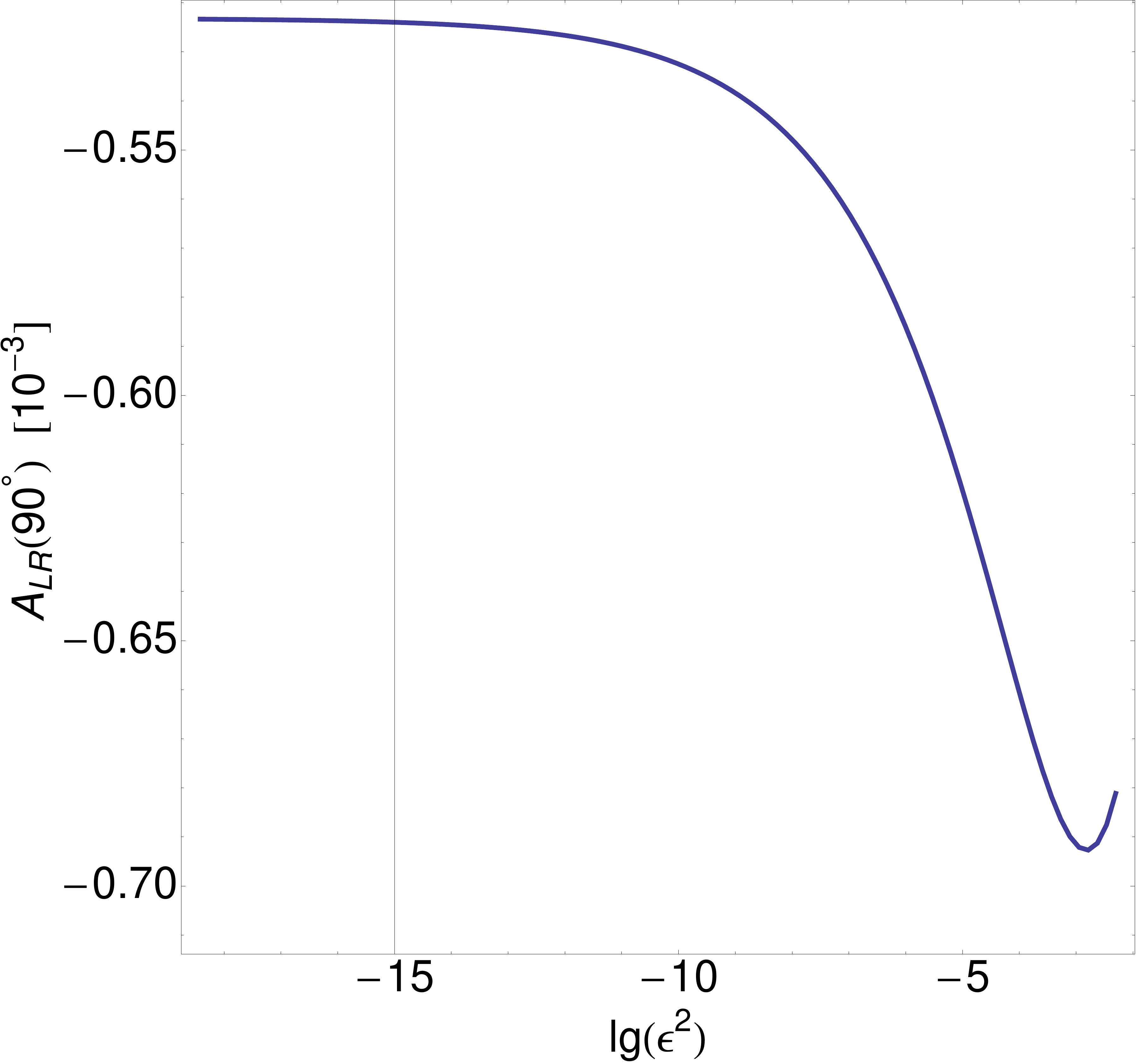} \includegraphics[scale=0.212]{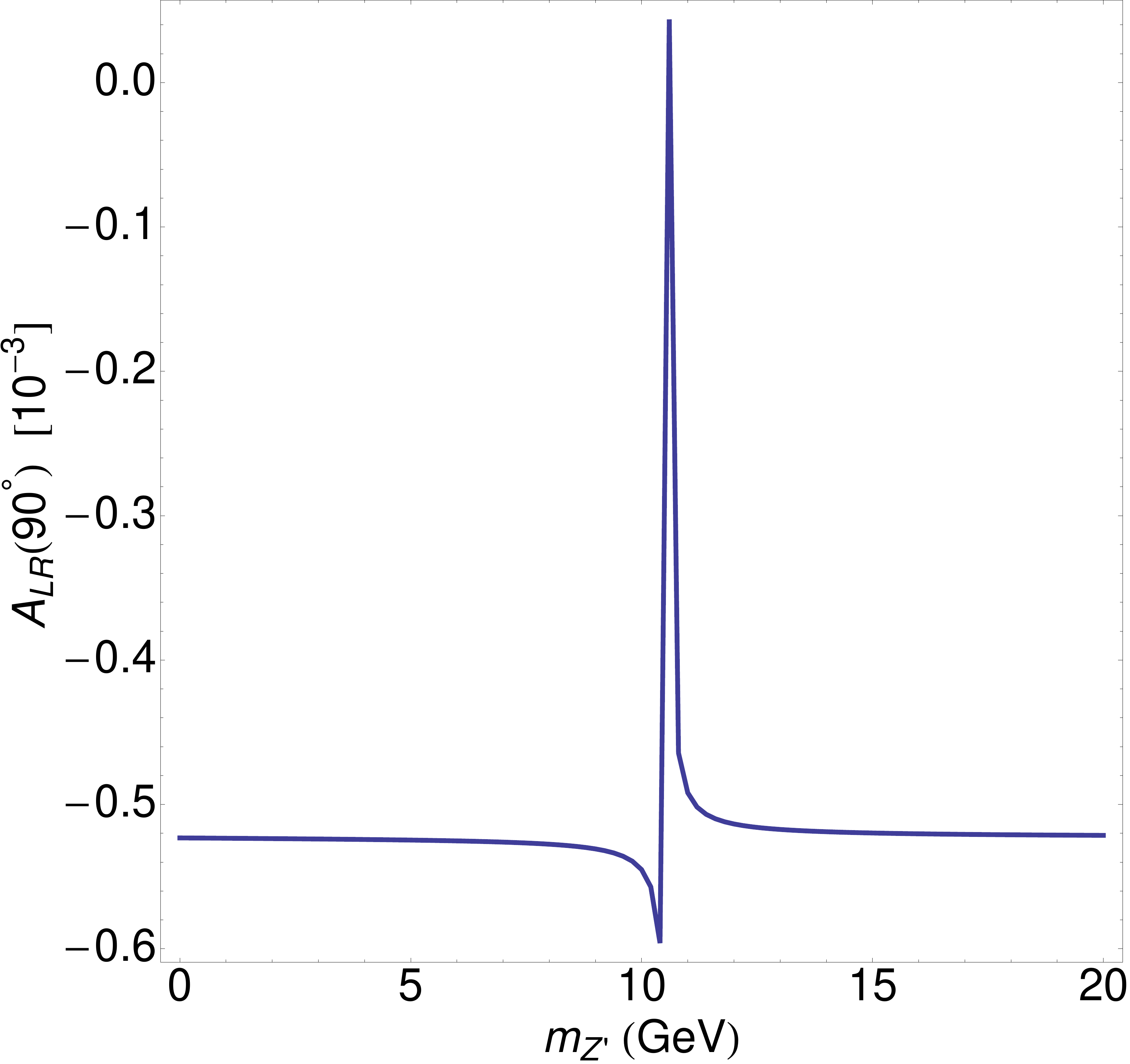}
\includegraphics[scale=0.21]{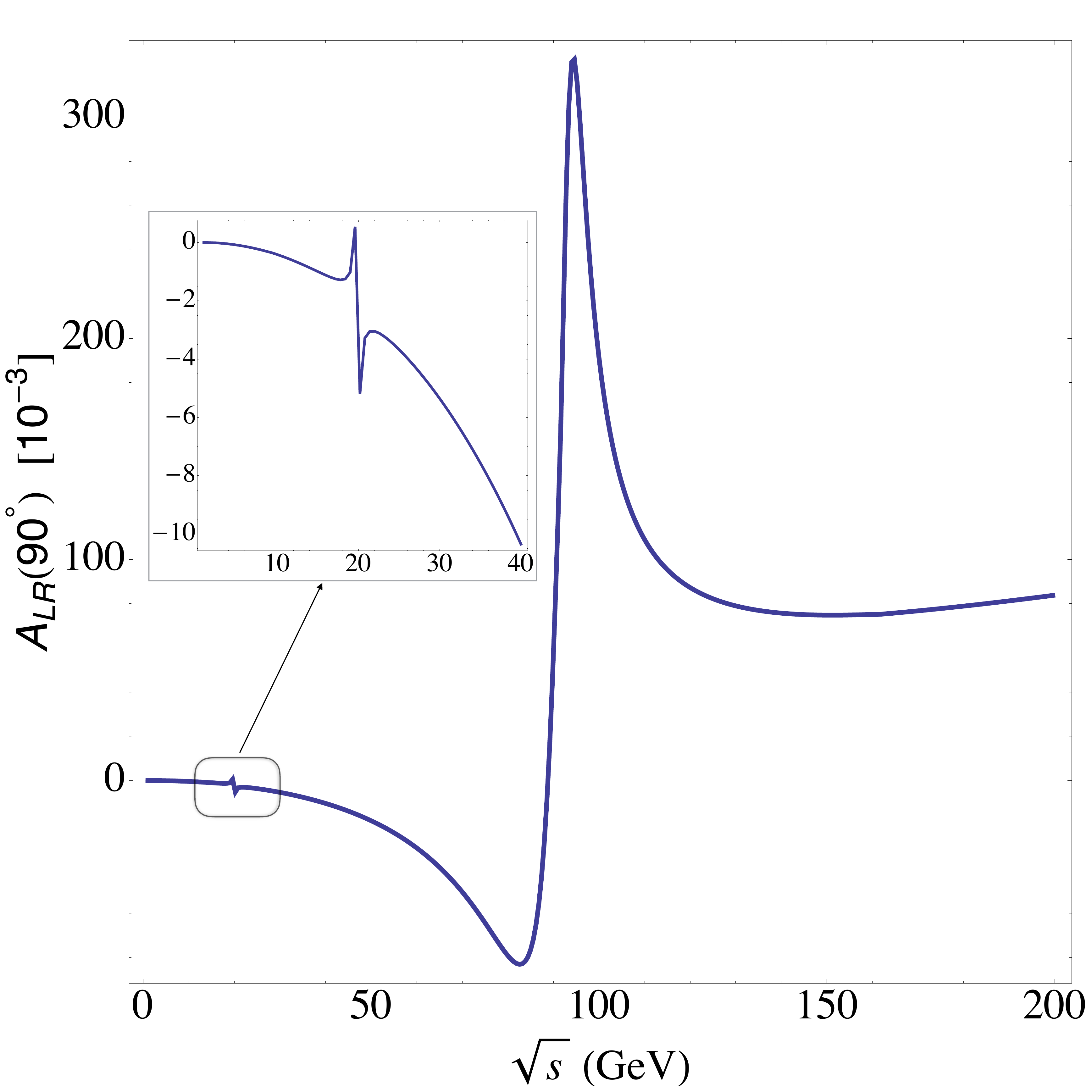}
\par\end{centering}

\caption{Figures show dependence of Belle II polarization asymmetry on mixing parameter 
$\epsilon^2$ (top left, for $m_{Z'}=8 \ {\rm GeV}$), $m_{Z'}$ (top righ, for $\epsilon^2=10^{-5}$)
and $\sqrt{s}$ (bottom center, for  $m_{Z'}=20 \ {\rm GeV}$ and $\epsilon^2=10^{-2}$ ). 
We choose for all graphs: $\delta^2=3\cdot10^{-5}$.}

\label{fig4}
\end{figure}

In Fig.\ref{fig3}, we explore a scenario of the small mixing $\epsilon$
and small $Z'_{\mu}$ mass, for MOLLER specifically. We show up-to-1\% deviation from
the SM central prediction exclusion plots for $Z'_{\mu}$, which is included
at LO and NLO orders. Inclusion of the $Z'_{\mu}$ at NLO order systematically
increase exclusion region for $\epsilon$ for all masses of $Z'_{\mu}$
by about 25\%. While this increase is not substantial, it could become
an important factor in the determination of $Z'_{\mu}$ mass and coupling if $Z'_{\mu}$ is discovered. According 
to \cite{key-19,key-20}, if no $Z'_{\mu}$ is discovered, MOLLER will exclude the region where $Z'_{\mu}$ is used
to explain $(g-2)_{\mu}$ anomaly. For Belle II (right plot
of Fig.\ref{fig3}), we concentrate on the resonance region at $\sqrt{s}=10.57\ {\rm GeV}$, 
where sensitivity of Belle II to the $Z'_{\mu}$ (up to NLO order) 
is the highest and is complimentary to the MOLLER experiment. 
For the $e^+ + e^- \rightarrow \mu^+ +\mu^-$ process, we also study the dependence of the asymmetry on the kinetic  
mixing parameter $\epsilon^2$ and $m_{Z'}$, which is shown in Fig.\ref{fig4} (top two plots). 
If we take mass of $Z'$ ($m_{Z'}=8\ {\rm GeV}$), 
close to the $\sqrt{s}=10.57\ {\rm GeV}$, the sensitivity of the asymmetry is the highest 
for $10^{-8} < \epsilon^2 < 10^{-4}$. In case of the fixed value for kinetic mixing,
$\epsilon^2=10^{-5}$, the sensitivity of the asymmetry to the variations of $m_{Z'}$ is very weak except for the narrow region of resonance around $\sqrt{s}=10.57\ {\rm GeV}$.  
This is because the leading $m_{Z'}$ dependence in asymmetry is determined by the ratio between kinetic mixing term $\epsilon^2$ and denominator of 
$Z'$ propagator $(\{t,s\}-m_{Z'}^2)$. 
In $t$-channel (MOLLER), at $ -t \rightarrow 0\ {\rm GeV}^2$, the asymmetry is proportional to $\epsilon^2/m_{Z'}^2$, 
and for light $Z'_{\mu}$-boson it has a rather high sensitivity to $m_{Z'}$. In s-channel (Belle II), 
we get $A_{LR}\propto\epsilon^2/(s-m_{Z'}^2)$, and for the $m^2_{Z'}/s \ll 1$ (light $Z'$) the sensitivity is rather low. 
If we take   $m^2_{Z'}/s \gg 1$ (heavy $Z'$), an overall contribution to the asymmetry due to $Z'$ is suppressed 
so the effect of new physics becomes negligible. As a result, for Belle II, if mass of $Z'_\mu$ is around $\sqrt{s}=10.57\ {\rm GeV}$, 
the sensitivity of that experiment 
to $m_{Z'}$ will be substantial. In Fig.\ref{fig4} (bottom, center), we show the overall dependence of asymmetry on the centre-of-mass energy. 
Here, we choose $m_{Z'}=20\ {\rm GeV}$ and $\epsilon^2=10^{-2}$, and it is evident that $Z'$-peak (if compared to $Z$-peak) is suppressed, which is due to the kinetic mixing parameter $\epsilon^2$. 
The $Z'$-peak is relatively small, but, with the precision proposed by Belle II, it should be clearly detectable. 

We conclude that the inclusion of NLO and NNLO electroweak radiative 
corrections is essential for the search of new physics at the precision frontier, 
and that the computer-algebra tools are indispensable for this task.

\section*{Acknowledgments}
This work has been supported by the Natural Sciences and Engineering Research Council of Canada (NSERC).

\end{document}